\documentclass[aps,pre,onecolumn,floatfix]{revtex4}

\usepackage{graphicx}
\usepackage{amssymb,amsfonts,amsmath}
\usepackage{epsf}
\usepackage{subfigure}
\usepackage{epstopdf}
\usepackage{mathrsfs}

\newcommand{\be}{\begin{equation}}
\newcommand{\ee}{\end{equation}}
\newcommand{\bea}{\begin{eqnarray}}
\newcommand{\eea}{\end{eqnarray}}

\begin{document}

\title{\centerline{\it\large In memoriam Christian Van den Broeck}
\vskip 0.8 cm
Microreversibility and driven Brownian motion \\ with hydrodynamic long-time correlations}

\author{Pierre Gaspard}
\affiliation{Center for Nonlinear Phenomena and Complex Systems, Universit{\'e} Libre de Bruxelles (U.L.B.), Code Postal 231, Campus Plaine, B-1050 Brussels, Belgium}

\begin{abstract}
A nonequilibrium fluctuation theorem is established for a colloidal particle driven by an external force within the hydrodynamic theory of Brownian motion, describing hydrodynamic memory effects such as the $t^{-3/2}$ power-law decay of the velocity autocorrelation function.  The generalized Langevin equation is obtained for the general case of slip boundary conditions between the particle and the fluid.  The Gaussian probability distributions for the particle to evolve in position-velocity space are deduced.  It is proved that the joint probability distributions of forward and time-reversed paths have a ratio depending only on the work performed by the external force and the fluid temperature, in spite of the nonMarkovian character of the generalized Langevin process.
\end{abstract}

\maketitle

\section{Introduction}

The manifestation of hydrodynamic memory effects in Brownian motion has been anticipated and predicted for long by theory \cite{M02,L21,VT45,DKS94,AW70,ZB70,HM73,D74,CS92,BF05}, and more recently observed in experiments using photon correlation dynamic laser light scattering \cite{PP81} and direct particle tracking \cite{JLKFF08,FGBMFFJ11,KSMLR14}.  In this regard, Langevin stochastic theory~\cite{L08} should be modified to include the memory effects induced by hydrodynamic flows around the colloidal particle.  Remarkably, generalized Langevin equations describing these effects can be deduced from fluctuating hydrodynamics since the seventies \cite{ZB70,HM73,D74,BM74,ABM75,BAM77,F76a,F76b,F77}.  Accordingly, the velocity autocorrelation function of the colloidal particle has a $t^{-3/2}$ power-law decay, referred to as hydrodynamic long-time tail.  The resulting generalized Langevin process is thus nonMarkovian.  However, the time evolution of the colloidal particle from initial to final conditions in position-velocity space is still described by a Gaussian probability distribution~\cite{D74}, as in the standard Langevin process~\cite{L08,K40,C43}.

Furthermore, interest has raised for Brownian motion driven by external forces into nonequilibrium regimes.  In particular, a fluctuation theorem has been proved for the Markovian Langevin process with an external force generating a nonequilibrium random drift~\cite{K98} and experiments have validated the nonequilibrium work fluctuation theorem~\cite{C99,J11} for colloidal particles driven by optical traps and described by standard Langevin equation~\cite{BSHSB06,CGP13}.  In this context, an important issue is to understand the effects of hydrodynamic long-time correlations upon nonequilibrium driving.  For these effects to manifest themselves, the mass of the displaced fluid should be comparable to the mass of the colloidal particle as long anticipated by Lorentz~\cite{M02,L21}.

The purpose of the present paper is to investigate the hydrodynamics long-time correlations for driven Brownian motion.  It is shown that the fluctuation theorem continues to hold exactly in spite of the nonMarkovian character of the generalized Langevin process describing the hydrodynamic effects.  We notice that fluctuation relations are already known for several types of nonMarkovian processes \cite{ZBCK05,SS07,MD07,EL08,AG08JSM,PV09,ABC10,MSVvW13,CK09,CLK12,DKC15}.  Here, our purpose is to consider the specific stochastic process that is deduced from the hydrodynamic theory of Brownian motion.  This latter is summarized in Sec.~\ref{sec:hydro}.  In Sec.~\ref{sec:GLP}, the Gaussian probability distributions ruling the time evolution of the colloidal particle in its phase space are obtained from the generalized Langevin equation.  The fluctuation theorem is presented in Sec.~\ref{sec:FT} and proved in App.~\ref{appA}.  Section~\ref{sec:expl} is devoted to the detailed calculation of the velocity autocorrelation function and its properties in the general case of slip boundary conditions.  The long-time properties of the random drift under the effect of an external force are given in Sec.~\ref{sec:EP}.  Concluding remarks are drawn in Sec.~\ref{sec:concl}.

\section{Hydrodynamic theory of Brownian motion}
\label{sec:hydro}

A colloidal particle of spherical shape, radius $a$, and mass density $\rho_{\rm s}$ is immersed in an incompressible fluid of mass density $\rho_{\rm f}$.  Because of the ceaseless thermal movements of the molecules composing the fluid, the velocity field ${\bf v}({\bf r},t)$ undergoes fluctuations, even if its mean value is vanishing at large distances from the particle, $\lim_{\Vert{\bf r}\Vert\to\infty}\langle{\bf v}({\bf r},t)\rangle=0$.  Since the fluid exerts a pressure on the surface of the colloidal particle, this latter is subjected to a fluctuating force at the origin of its Brownian motion.  For this reason, the stochastic equation describing Brownian motion can be deduced from fluctuating hydrodynamics \cite{LL80Part2,OS06}, as shown in the seventies \cite{ZB70,HM73,D74,BM74,ABM75,BAM77,F76a,F76b,F77}.  We notice that the effects of compressibility should manifest themselves at frequencies $\omega\gg v_{\rm sound}/a\sim 10^9$~s$^{-1}$ \cite{BM74c}, so that they play a negligible role for Brownian motion in a fluid at rest.

\subsection{Interaction between the particle and the fluid}

For our purpose, the equations of fluctuating hydrodynamics are solved to obtain the velocity field around the colloidal particle including the contribution of the fluctuating velocity field.  Since the velocity field determines the pressure tensor ${\boldsymbol{\mathsf P}}$ in the fluid, the stochastic differential equation for the colloidal particle velocity ${\bf V}(t)$ is obtained from Newton's equation
\be
m_{\rm s} \frac{d{\bf V}(t)}{dt} = -\int_{{\cal S}(t)} {\boldsymbol{\mathsf P}}({\bf r},t)\cdot{\bf n} \, d\Sigma + {\bf F}_{\rm ext} \, ,
\label{Langevin-Eq-0}
\ee
by integrating the fluid pressure tensor over the moving surface ${\cal S}(t)$ of the colloidal particle \cite{BM74,MB74}, where $m_{\rm s}=(4\pi a^3/3)\rho_{\rm s}$ is the mass of the particle and ${\bf F}_{\rm ext}$ is a possible external force.  If the system evolves in a gravitational field of acceleration $\bf g$, the external force is given by ${\bf F}_{\rm ext}=(m_{\rm s}-m_{\rm f}){\bf g}$, where $m_{\rm f}=(4\pi a^3/3)\rho_{\rm f}$ is the mass of the fluid displaced by the particle, according to Archimedes' principle.

In the framework of fluctuating hydrodynamics \cite{LL80Part2,OS06},  the velocity field of an incompressible fluid obeys $\pmb{\nabla}\cdot{\bf v}=0$ and the fluctuating Navier-Stokes equations expressed as
\be\label{NS-eqs}
\rho_{\rm f}\left(\partial_t{\bf v} + {\bf v}\cdot\pmb{\nabla}{\bf v}\right) = -\pmb{\nabla}\cdot{\boldsymbol{\mathsf P}} + \pmb{f}_{\rm ext}\, ,
\ee
in terms of a possible external force density $\pmb{f}_{\rm ext}$ and the pressure tensor 
\be\label{pressure}
{\boldsymbol{\mathsf P}}= P\, {\boldsymbol{\mathsf 1}} - \eta \left(\pmb{\nabla}{\bf v} + \pmb{\nabla}{\bf v}^{\rm T}\right) +{\boldsymbol{\mathsf \pi}}
\ee
where $P$ is the hydrostatic pressure, ${\boldsymbol{\mathsf 1}}$ is the $3\times 3$ unit matrix, $\eta$ is the shear viscosity of the fluid, the superscript~T denotes the transpose, and ${\boldsymbol{\mathsf\pi}}=(\pi_{ij})$ is a fluctuating pressure tensor with components given by Gaussian white noise fields satisfying
\be \label{pi-correl}
\langle \pi_{ij}({\bf r},t)\rangle = 0 \qquad\mbox{and}\qquad\langle \pi_{ij}({\bf r},t)\, \pi_{kl}({\bf r}',t')\rangle = 2 k_{\rm B}T \eta \left(\delta_{ik}\delta_{jl}+\delta_{il}\delta_{jk}\right)\delta({\bf r}-{\bf r}') \, \delta(t-t')\, ,
\ee
$k_{\rm B}$ being Boltzmann's constant and $T$ the temperature.

Since the colloidal particle is solid, the velocity field is given for $\Vert{\bf r}-{\bf R}(t)\Vert <a$ by
\be\label{v-sol}
{\bf v}({\bf r},t) = {\bf V}(t) + \pmb{\Omega}(t)\times \left[ {\bf r}-{\bf R}(t)\right] ,
\ee
where ${\bf R}(t)$ denotes the mass center of the particle, ${\bf V}(t)=d{\bf R}(t)/dt$ its velocity, and $\pmb{\Omega}(t)$ its angular velocity \cite{MB74}.

At the interface $\Vert{\bf r}-{\bf R}(t)\Vert=a$ between the solid particle and the fluid, the following boundary conditions are considered \cite{BM74,ABM75,BAM77,F76a,F76b,F77}.  On the one hand, the fluid velocity field in the direction $\bf n$ normal to the interface should satisfy
\be
{\bf n}\cdot{\bf v}({\bf r},t) = {\bf n}\cdot{\bf V}(t) \, .
\ee
On the other hand, in the tangential directions ${\boldsymbol{\mathsf 1}}_{\bot}={\boldsymbol{\mathsf 1}}-{\bf n}{\bf n}$, the boundary conditions on the velocity field are expressed as \cite{BAM77}
\be
\lambda \, {\boldsymbol{\mathsf 1}}_{\bot}\cdot\left\{ {\bf v}({\bf r},t) -{\bf V}(t) - \pmb{\Omega}(t)\times\left[{\bf r}-{\bf R}(t)\right]\right\} 
={\boldsymbol{\mathsf 1}}_{\bot}\cdot\left\{  \eta \left[\pmb{\nabla}{\bf v}({\bf r},t) +\pmb{\nabla}{\bf v}({\bf r},t)^{\rm T}\right]\cdot{\bf n} +\pmb{f}^{\rm s}_{\rm fl}({\bf r},t)\right\} ,\label{v-bc2}
\ee
in terms of the sliding friction coefficient $\lambda$ and the interfacial Gaussian white noise process satisfying 
\be
\langle\pmb{f}_{\rm fl}^{\rm s}({\bf r},t)\rangle = 0 \qquad \mbox{and} \qquad \delta^{\rm s}({\bf r},t)\, \langle\pmb{f}_{\rm fl}^{\rm s}({\bf r},t)\, \pmb{f}_{\rm fl}^{\rm s}({\bf r}',t')\rangle\, \delta^{\rm s}({\bf r}',t') = 2 k_{\rm B}T \lambda \, \delta^{\rm s}({\bf r},t) \, \delta({\bf r}-{\bf r}') \, \delta(t-t') \, {\boldsymbol{\mathsf 1}}_{\bot} \, ,
\label{f-f}
\ee
where $\delta^{\rm s}({\bf r},t)$ is the interfacial Dirac distribution introduced in Ref.~\cite{BAM76}.  We note that sliding friction causes the slippage of the velocity field at the interface with the slip length
\be
b\equiv \frac{\eta}{\lambda}\, .
\label{b}
\ee

\subsection{Solution of the problem}

At low Reynolds numbers, the Navier-Stokes equations and the boundary conditions can be linearized and solved by taking Fourier transforms according to
\be
\phi(\omega) \equiv \int_{-\infty}^{+\infty}  {\rm e}^{i\omega t} \, \phi(t) \, dt
\ee
for any function $\phi(t)$ of time.  Using the method of induced force~\cite{BM74,ABM75,BAM77,F76a,F76b,F77}, the following equation is obtained
\be
-i\omega m_{\rm s} {\bf V}(\omega) = -\zeta(\omega) \, {\bf V}(\omega) + {\bf F}_{\rm fl}(\omega) + 2\pi \, \delta(\omega) \, {\bf F}_{\rm ext} 
\label{eq-w-1}
\ee
in terms of the frequency-dependent friction coefficient
\be
\zeta(\omega) = 6 \pi \eta a \biggl[\frac{(1+2b/a)(1+a \alpha)}{(1+3b/a)+b \alpha} +\frac{a^2\alpha^2}{9}  \biggr] \qquad\mbox{with} \qquad \alpha=\sqrt{-i\omega/\nu}
\label{zeta}
\ee
(where $\nu\equiv\eta/\rho_{\rm f}$ is the kinematic viscosity) and the fluctuating force ${\bf F}_{\rm fl}(\omega)$ obeying the fluctuation-dissipation theorem
\be
\langle {\bf F}_{\rm fl}(\omega)\rangle = 0 \qquad\mbox{and}  \qquad
\langle {\bf F}_{\rm fl}(\omega)\, {\bf F}_{\rm fl}^*(\omega')\rangle = 4\pi\, k_{\rm B}T \, {\rm Re}\, \zeta(\omega) \, \delta(\omega-\omega') \, {\boldsymbol{\mathsf 1}} \, .
\ee
A recent survey of this calculation is given in Ref.~\cite{GK18a}.

Expliciting the last term going as $\alpha^2$ in Eq.~(\ref{zeta}), the frequency-dependent friction coefficient
can be rewritten as
\be
\zeta(\omega) = \zeta_{\rm d}(\omega) - i \omega \, \frac{m_{\rm f}}{2}
\label{zeta2}
\ee
in terms of the mass $m_{\rm f}$ of the fluid displaced.  Moreover, we have that
\be
{\rm Re}\, \zeta(\omega) = {\rm Re}\, \zeta_{\rm d}(\omega) \, ,
\ee
so that the last term of Eq.~(\ref{zeta2}) does not contribute to damping, the effects of which are only described by the contribution
\be
\zeta_{\rm d}(\omega) = \gamma\,  \frac{1+a \sqrt{-i\omega/\nu}}{1+B \sqrt{-i\omega/\nu}} \, ,
\label{zeta-d}
\ee
where
\be
\gamma\equiv \zeta(0)= \zeta_{\rm d}(0)= 6\pi\eta a \, \frac{1+2b/a}{1+3b/a}
 \label{friction}
\ee
is the low-frequency limit of the friction coefficient and
\be
B \equiv \frac{b}{1+3b/a} \, .
\label{big-B}
\ee
Furthermore, the last term in Eq.~(\ref{zeta2}) gives a contribution in Eq.~(\ref{eq-w-1}) similar to the inertial term in the left-hand side of that equation.  Therefore, these terms can be gathered to get
\be
-i\omega m\,  {\bf V}(\omega) = -\zeta_{\rm d}(\omega) \, {\bf V}(\omega) + {\bf F}_{\rm fl}(\omega) + 2\pi \, \delta(\omega) \, {\bf F}_{\rm ext} 
\label{eq-w-2}
\ee
with the total mass
\be
m \equiv m_{\rm s} + \frac{m_{\rm f}}{2} \, ,
\label{tot-m}
\ee
including the mass of the solid particle itself together with half the mass of the fluid displaced, as shown in Refs.~\cite{CL56,MR83,D74}.  

Because of the frequency dependence of the friction coefficient $\zeta_{\rm d}(\omega)$, the stochastic equation of motion corresponding to Eq.~(\ref{eq-w-2}) by inverse Fourier transform should have the form of a generalized Langevin equation, as discussed here below.

\section{Generalized Langevin process}
\label{sec:GLP}

\subsection{Generalized Langevin equation}

According to the discussion of previous section following the results of Refs.~\cite{HM73,D74,BM74,ABM75,BAM77,F76a,F76b,F77}, the stochastic motion of the Brownian particle should be described by a generalized Langevin equation of the form
\be
m \, \frac{d{\bf V}(t)}{dt} = -\int_0^t \Gamma(t-t') \, {\bf V}(t') \, dt' + {\bf F}_{\rm fl}(t) + {\bf F}_{\rm ext}
\label{eq-V}
\ee
with a memory kernel $\Gamma(t-t')$ determined by the frequency-dependent friction coefficient~(\ref{zeta-d}), and a fluctuating force given by a Gaussian white noise characterized by
\be
\langle {\bf F}_{\rm fl}(t)\rangle = 0 \qquad\mbox{and}  \qquad
\langle {\bf F}_{\rm fl}(t)\, {\bf F}_{\rm fl}(t')\rangle = k_{\rm B}T \, \Gamma(\vert t-t'\vert) \, {\boldsymbol{\mathsf 1}} \, . \label{F-autocorrel}
\ee
Moreover, the fluctuating force is not correlated with the initial condition of the velocity, $\langle {\bf V}(0)\,{\bf F}_{\rm fl}(t)\rangle = 0$ for $t>0$ \cite{D74,B79}.  Accordingly, the stochastic process considered is Gaussian, but not Markovian.  

\subsection{General solution}

Since Eq.~(\ref{eq-V}) is linear, it can be solved by direct integration, as shown in Ref.~\cite{D74}.  Next, the position is found by integrating $d{\bf R}(t)/dt={\bf V}(t)$.  Introducing the function $K(t)$ satisfying
\be
m \, \frac{dK(t)}{dt} = -\int_0^t \Gamma(t-t') \, K(t') \, dt' \qquad \mbox{with} \qquad K(0)=1 \, ,
\label{eq-K}
\ee
the particle velocity and position are given by
\bea
&&{\bf V}(t) = K(t) \, {\bf V}_0 + \frac{1}{m} \int_0^t dt' \, K(t-t') \left[ {\bf F}_{\rm fl}(t') + {\bf F}_{\rm ext}\right] , \\
&&{\bf R}(t) = {\bf R}_0 + \int_0^t dt' \, K(t') \, {\bf V}_0 + \frac{1}{m} \int_0^t dt' \, \int_0^{t'} dt'' \, K(t'-t'') \left[ {\bf F}_{\rm fl}(t'') + {\bf F}_{\rm ext}\right] ,
\eea
 in terms of the initial velocity and position, ${\bf V}_0$ and ${\bf R}_0$.
 
 Beside the function $K(t)$ introduced with Eq.~(\ref{eq-K}), we may define two further functions
 \bea
&&  L(t) \equiv \int_0^t K(t') \, dt' \, , \label{L}\\
&&  M(t) \equiv \int_0^t L(t') \, dt' \, , \label{M}
\eea
 which will also play an important role in the following.
 
 \subsection{Conditional and joint probability densities}
 
 The stochastic process described by the generalized Langevin equation has Gaussian conditional and joint probability densities to move from the initial condition $({\bf R}_0,{\bf V}_0)$ at time $t=0$ to the phase-space point $({\bf R},{\bf V})$ at time~$t$.  As shown in Refs.~\cite{C43,D74}, the conditional probability density is given by
 \be
 p({\bf R},{\bf V},t\vert{\bf R}_0,{\bf V}_0,0) = \langle\delta\left[{\bf X}-{\bf x}(t)\right]\, \delta\left[{\bf Y}-{\bf y}(t)\right]\rangle 
 \ee
in terms of the variables
\bea
&&{\bf X} \equiv {\bf R} - {\bf R}_0 - L(t) \, {\bf V}_0 - M(t) \, \frac{{\bf F}_{\rm ext}}{m} \, , \label{X}\\
&&{\bf Y} \equiv {\bf V} - K(t) \, {\bf V}_0 - L(t) \, \frac{{\bf F}_{\rm ext}}{m}\, , \label{Y}
\eea
and the Gaussian white noise processes
\bea
&&{\bf x}(t) \equiv \frac{1}{m} \int_0^t dt' \, \int_0^{t'} dt'' \, K(t'-t'') \, {\bf F}_{\rm fl}(t'')\, , \\
&&{\bf y}(t) \equiv \frac{1}{m} \int_0^t dt' \, K(t-t')\, {\bf F}_{\rm fl}(t')\, .
\eea
Consequently, the conditional probability density is found to be given by
 \be
p({\bf R},{\bf V},t\vert{\bf R}_0,{\bf V}_0,0) = \frac{1}{(2\pi)^3\sqrt{\det{\boldsymbol{\mathsf C}}}} \, \exp\left(-\frac{1}{2}\, {\bf Z}^{\rm T}\cdot{\boldsymbol{\mathsf C}}^{-1}\cdot{\bf Z}\right)
\ee
with
\be
{\bf Z} = 
\left(
\begin{array}{c}
{\bf X} \\
{\bf Y} \\
\end{array}
\right)
\ee
and the correlation matrix
\be
{\boldsymbol{\mathsf C}} = 
\left[
\begin{array}{cc}
\langle{\bf x}(t){\bf x}(t)\rangle & \langle{\bf x}(t){\bf y}(t)\rangle \\
\langle{\bf y}(t){\bf x}(t)\rangle & \langle{\bf y}(t){\bf y}(t)\rangle \\
\end{array}
\right] = 
\left[
\begin{array}{cc}
F(t) & H(t) \\
H(t) & G(t) \\
\end{array}
\right] \, {\boldsymbol{\mathsf 1}}\, .
\ee
The elements of this matrix can be directly calculated to get~\cite{C43,D74}
\bea
&& F(t) = \frac{1}{3} \, \langle{\bf x}(t)^2\rangle =  \frac{k_{\rm B}T}{m} \left[ 2\, M(t) - L(t)^2\right]  , \label{F}\\
&& G(t) = \frac{1}{3} \, \langle{\bf y}(t)^2\rangle =  \frac{k_{\rm B}T}{m} \left[ 1 - K(t)^2\right]  , \label{G}\\
&& H(t) = \frac{1}{3} \, \langle{\bf x}(t)\cdot{\bf y}(t)\rangle =  \frac{k_{\rm B}T}{m} L(t) \, \left[ 1 - K(t)\right]  , \label{H}
\eea
in terms of the previously introduced functions~(\ref{eq-K}), (\ref{L}), and~(\ref{M}).  Since ${\bf y}=d{\bf x}/dt$, we have that $dF/dt=2H$, which can be integrated once $H(t)$ is known to get $F(t)$.  Therefore, the conditional probability density has the following Gaussian form~\cite{C43,D74}
\be
p({\bf R},{\bf V},t\vert{\bf R}_0,{\bf V}_0,0) = \frac{1}{8\pi^3(FG-H^2)^{3/2}} \, \exp\left[-\frac{G{\bf X}^2-2H{\bf X}\cdot{\bf Y} +F{\bf Y}^2}{2(FG-H^2)}\right] .
\ee

The distribution of the velocity is obtained by integrating over the position $\bf R$ to find
\be
{\mathscr P}({\bf V},t\vert{\bf V}_0,0) = \int p({\bf R},{\bf V},t\vert{\bf R}_0,{\bf V}_0,0)\, d{\bf R} = \frac{1}{(2\pi G)^{3/2}} \, \exp\left(-\frac{{\bf Y}^2}{2G}\right) ,
\label{p-V}
\ee
which converges in the long-time limit towards the Maxwell-Boltzmann equilibrium probability distribution of density
\be
p_{\rm eq}({\bf V}) \equiv \lim_{t\to\infty} {\mathscr P}({\bf V},t\vert{\bf V}_0,0) = \left( \frac{m}{2\pi k_{\rm B}T}\right)^{3/2} \, \exp\left(-\frac{m{\bf V}^2}{2k_{\rm B}T}\right) ,
\label{MB}
\ee
corresponding to the kinetic energy of the total mass~(\ref{tot-m}). Thermalization thus happens for the colloidal particle of mass $m_{\rm s}$ accompanied with the mass $m_{\rm f}/2$ of fluid, because of the inertial effect induced by hydrodynamics \cite{D74}.

The joint probability density to move from the position ${\bf R}_0$ and the velocity ${\bf V}_0$ distributed according to the equilibrium Maxwell-Boltzmann density~(\ref{MB}) at time $t=0$ to the phase-space point $({\bf R},{\bf V})$ at time~$t$ is defined as
 \be
p({\bf R},{\bf V},t;{\bf R}_0,{\bf V}_0,0) \equiv p({\bf R},{\bf V},t\vert{\bf R}_0,{\bf V}_0,0) \, p_{\rm eq}({\bf V}_0)\, .
\label{joint-p}
\ee

Under these conditions, the conditional probability density to move from the initial position ${\bf R}_0$ towards the position $\bf R$ at time~t is given by
\be
{\cal P}({\bf R},t\vert{\bf R}_0,0) \equiv \int p({\bf R},{\bf V},t;{\bf R}_0,{\bf V}_0,0) \, d{\bf V} \, d{\bf V}_0
= \left[ \frac{\beta\, m}{4\pi M(t)}\right]^{3/2} \, \exp\left\{ -\frac{\beta\, m}{4\, M(t)}\left[{\bf R}-{\bf R}_0 - M(t)  \frac{{\bf F}_{\rm ext}}{m}\right]^2\right\} ,
\label{p-R}
\ee
where $\beta=(k_{\rm B}T)^{-1}$. 

Furthermore, the velocity autocorrelation function has the form
\be
\langle \Delta{\bf V}(0) \cdot\Delta{\bf V}(t)\rangle = 3 \, \frac{k_{\rm B}T}{m} \, K(t) \, ,
\label{V-correl}
\ee
where $\Delta{\bf V}(t)={\bf V}(t)-\langle{\bf V}(t)\rangle$.

\section{General fluctuation theorem}
\label{sec:FT}

In order to investigate the consequences of microreversibility on the generalized Langevin process, we compare two different paths that are mapped onto each other by time reversal, which consists in exchanging the initial and final positions while reversing the velocities.  

On the one hand, the forward path
\be
({\bf R}_0,{\bf V}_0,0)\ {\overset{\rm F}\longrightarrow} \ ({\bf R},{\bf V},t)
\ee
has the joint probability density
\be
p_{\rm F} \equiv p({\bf R},{\bf V},t;{\bf R}_0,{\bf V}_0,0) 
\label{pF}
\ee
defined by Eq.~(\ref{joint-p}).

On the other hand, the reversed path is given by
\be
({\bf R},-{\bf V},0) \ {\overset{\rm R}\longrightarrow} \ ({\bf R}_0,-{\bf V}_0,t)
\ee
having the joint probability density
\be
p_{\rm R} \equiv p({\bf R}_0,-{\bf V}_0,t;{\bf R},-{\bf V},0) \, .
\label{pR}
\ee

\begin{figure}[h]
\centerline{\scalebox{0.5}{\includegraphics{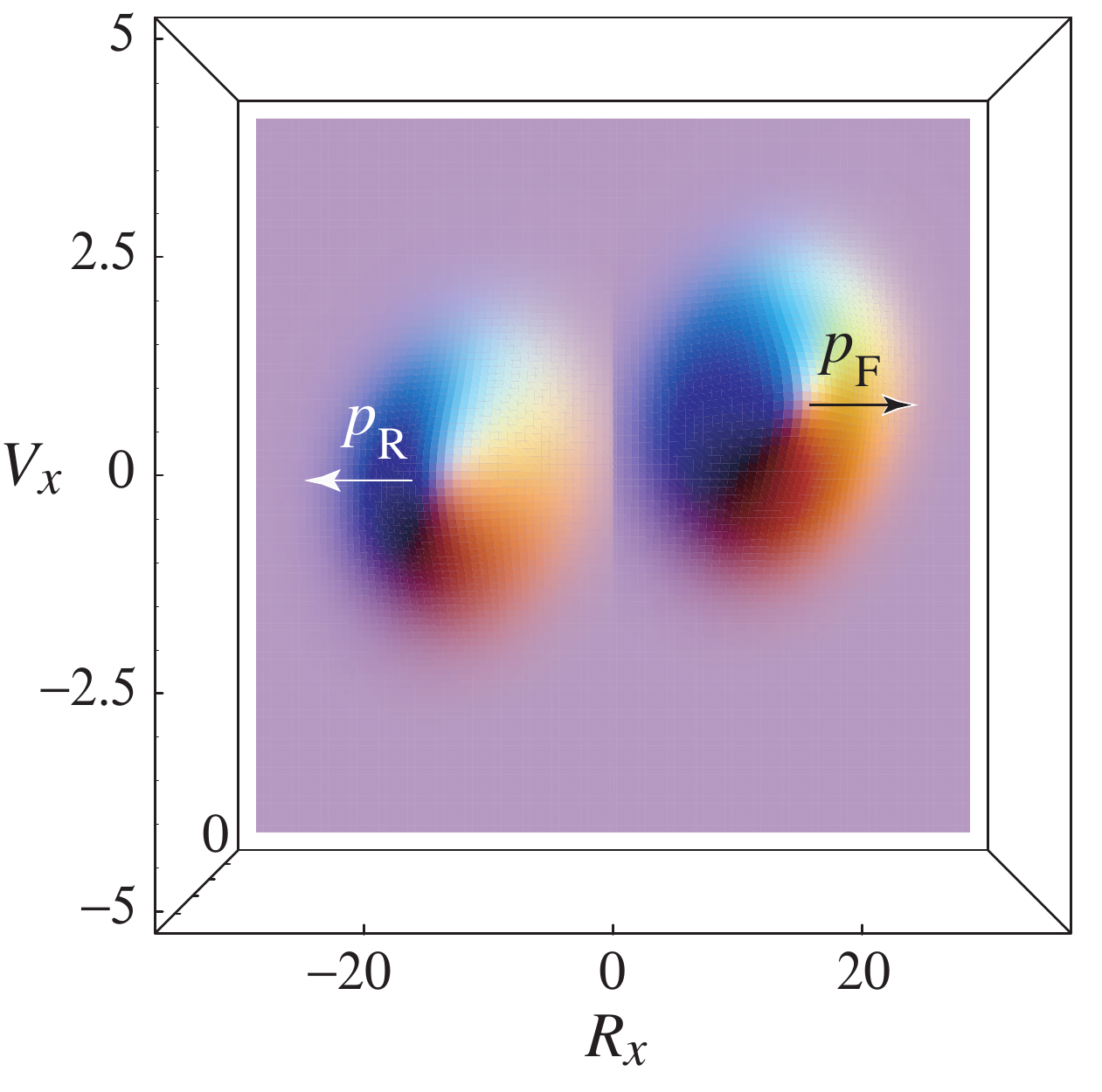}}}
\caption{Schematic one-dimensional representation of the joint probability densities~(\ref{pF}) and~(\ref{pR}) of the forward and reversed paths in the phase space $(R_x,V_x)$.  The probability density of the forward path moves in the direction of the external force $F_{x\; {\rm ext}}>0$ and the one of the reversed path in the opposite direction.}
\label{fig1}
\end{figure}

Figure~\ref{fig1} depicts schematically the joint probability densities~(\ref{pF}) and~(\ref{pR}) in the phase space $(R_x,V_x)$.  Under the effect of the external force, these probability densities are moving in opposite directions, so that their overlap rapidly decreases, as time increases.  

The remarkable result is that the following fluctuation theorem holds for any generalized Langevin process defined by Eqs.~(\ref{eq-V}) and~(\ref{F-autocorrel}).

\vskip 0.5 cm

\noindent{\bf Fluctuation theorem.}  {\it The joint probability densities~(\ref{pF}) and~(\ref{pR}) of the forward and reversed paths obey the relation}
\be
\frac{p({\bf R},{\bf V},t;{\bf R}_0,{\bf V}_0,0)}{p({\bf R}_0,-{\bf V}_0,t;{\bf R},-{\bf V},0)} = {\rm e}^{\beta {\bf F}_{\rm ext}\cdot({\bf R}-{\bf R}_0)} \, ,
\label{FT}
\ee
{\it for any functions~(\ref{F}), (\ref{G}), and~(\ref{H}) given in terms of the functions~(\ref{eq-K}), (\ref{L}), and~(\ref{M}) at every time $t\ge 0$.}

\vskip 0.5 cm

This theorem is proved in App.~\ref{appA}.  The ratio of the joint probability densities of the forward and reversed paths is thus independent of hydrodynamic effects.  If the external force vanishes, ${\bf F}_{\rm ext}=0$, we recover the principle of detailed balance, as required by microreversibility under equilibrium conditions.

Using the definition~(\ref{joint-p}) of the joint probability densities and the Maxwell-Boltzmann distribution~(\ref{MB}), the following corollary is obtained for the corresponding conditional probability densities
\be
\frac{p({\bf R},{\bf V},t\vert{\bf R}_0,{\bf V}_0,0)}{p({\bf R}_0,-{\bf V}_0,t\vert{\bf R},-{\bf V},0)} = {\rm e}^{\beta (W-\Delta E)} \, ,
\label{FT-c}
\ee
in terms of the work performed by the external force during the displacement of the particle from ${\bf R}_0$ to $\bf R$, and the change in the kinetic energy of the particle,
\be
W\equiv {\bf F}_{\rm ext}\cdot({\bf R}-{\bf R}_0) \qquad \mbox{and}\qquad \Delta E \equiv \frac{1}{2}  m  {\bf V}^2 - \frac{1}{2}  m  {\bf V}_0^2 \, ,
\ee
establishing the energy balance of stochastic energetics~\cite{S10}.

Moreover, the conditional probability densities in position space~(\ref{p-R}) satisfy the following fluctuation relation
\be
\frac{{\cal P}({\bf R},t\vert{\bf R}_0,0)}{{\cal P}({\bf R}_0,t\vert{\bf R},0)} = {\rm e}^{\beta {\bf F}_{\rm ext}\cdot({\bf R}-{\bf R}_0)} \, ,
\label{FT-R}
\ee
independently of the specificities of the generalized Langevin process.

\section{Hydrodynamic long-time correlations}
\label{sec:expl}

\subsection{Slip boundary conditions}

In order to obtain the function $K(t)$ determining the velocity autocorrelation function~(\ref{V-correl}), we solve Eq.~(\ref{eq-K}) by Laplace transform defined as
\be
\tilde\phi(z) \equiv \int_0^{\infty} {\rm e}^{-zt} \, \phi(t) \, dt 
\ee
for an arbitrary function $\phi(t)$ of time.  We thus have that the Laplace transform of the function $K(t)$ is given by
\be
\tilde K(z) = \left[ z + \frac{1}{m} \, \tilde\Gamma(z)\right]^{-1}
\label{K-tilde}
\ee
where the Laplace transform of the memory kernel $\Gamma(t)$ is related to the non-inertial part~(\ref{zeta-d}) of the frequency-dependent friction coefficient by the analytic continuation $z=-i\omega$, so that
\be
\tilde\Gamma(z) = \gamma\,  \frac{1+a \sqrt{z/\nu}}{1+B \sqrt{z/\nu}}
\label{Gamma-tilde}
\ee
with the low-frequency limit~(\ref{friction}) of the friction coefficient and the parameter~(\ref{big-B}).  Here, we introduce the thermalization time of the velocity distribution function
\be
\tau \equiv \frac{m}{\gamma} \, ,
\label{tau}
\ee
as well as the parameters
\be
\lambda\equiv \frac{\nu\tau}{a^2} \qquad\mbox{and}\qquad \mu\equiv \frac{\nu\tau}{B^2} \, .
\ee
The function $K(t)$ is obtained by the inverse Laplace transform of Eq.~(\ref{K-tilde}) with~Eq.~(\ref{Gamma-tilde})
\be
K(t) = \int_{c-i\infty}^{c+i\infty} \frac{dz}{2\pi i} \, {\rm e}^{zt} \, \tilde K(z) \, ,
\label{inv-Laplace}
\ee
where the constant $c$ should exceed the real part of all the singularities of $\tilde K(z)$.  This latter presents a branch cut on the negative half of the real axis in the complex plane of the variable $z$.  Deforming the integration contour around this branch cut and setting $z=-r/\tau$, we find the following integral representation for $t\ge 0$:
\be
K(t) = \frac{1}{\pi} \left( \frac{1}{\sqrt{\lambda}}-\frac{1}{\sqrt{\mu}}\right) \int_0^{\infty} \frac{\sqrt{r} \ {\rm e}^{-rt/\tau}}{(r-1)^2 + r\left(\frac{1}{\sqrt{\lambda}}-\frac{r}{\sqrt{\mu}}\right)^2} \, dr \, .
\label{K-int}
\ee

\begin{figure}[h]
\centerline{\scalebox{0.7}{\includegraphics{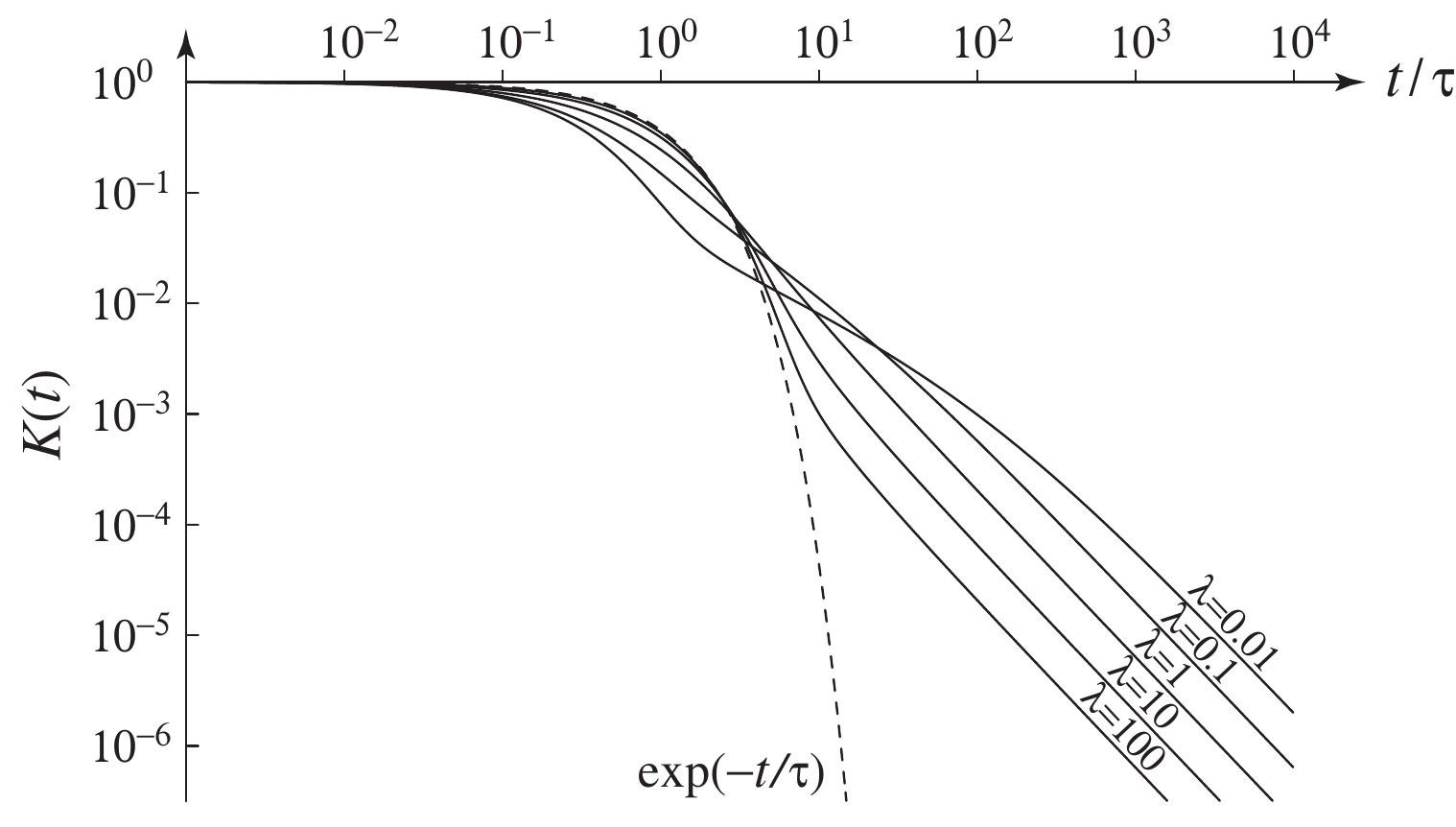}}}
\caption{The function~(\ref{K-int}) versus the rescaled time $t/\tau$ for the slip length $b=2a$ and the parameter values $\lambda=0.01$-$100$ (solid lines).  The other parameter is given by $\mu=3.5^2\lambda$ for this value of the slip length.  The dashed line shows the exponential decay expected for the standard Langevin process.}
\label{fig2}
\end{figure}

This function, as well as the functions~(\ref{L}) and~(\ref{M}), are analyzed in App.~\ref{appB}.  They behave as follows for long times $t\to\infty$:
\bea
&&K(t) \simeq \frac{1}{2\sqrt{\pi}} \left( \frac{1}{\sqrt{\lambda}}-\frac{1}{\sqrt{\mu}}\right)  \left( \frac{\tau}{t}\right)^{3/2} \, , \label{K-long}\\
&&L(t) \simeq \tau\left[ 1 -\frac{1}{\sqrt{\pi}} \left( \frac{1}{\sqrt{\lambda}}-\frac{1}{\sqrt{\mu}}\right)  \left( \frac{\tau}{t}\right)^{1/2}\right]  , \label{L-long}\\
&&M(t) \simeq \tau t \left[ 1 -\frac{2}{\sqrt{\pi}} \left( \frac{1}{\sqrt{\lambda}}-\frac{1}{\sqrt{\mu}}\right)  \left( \frac{\tau}{t}\right)^{1/2}\right]  . \label{M-long}
\eea
At short times $t\to 0$, the function $K(t)$ decreases from its initial condition $K(0)=1$ according to
\be
K(t) \simeq 1 - \frac{t}{\tau} \, \sqrt{\frac{\mu}{\lambda}} \, .
\label{K-short}
\ee

Figure~\ref{fig2} shows the function $K(t)$ for different values of the parameter $\lambda$.  We observe the long-time tail $K(t)\sim t^{-3/2}$ due to hydrodynamics, which generates the persistence of motion and the memory of the initial velocity.  At short times, the function is seen to converge to $K(0)=1$, according to Eq.~(\ref{K-short}).

\subsection{Stick boundary conditions}

The stick boundary conditions ($b=0$) correspond to the limit $\mu=\infty$.  In this case, the long-time behavior of the different functions is still given by Eqs.~(\ref{K-long})-(\ref{M-long}).  However, the function $K(t)$ behaves at short times as
\be
K(t) = 1 - 2 \sqrt{\frac{t}{\pi\lambda\tau}} +O(t) \, ,
\label{K-short-stick}
\ee
which is deduced in App.~\ref{appB}.

\begin{figure}[h]
\centerline{\scalebox{0.7}{\includegraphics{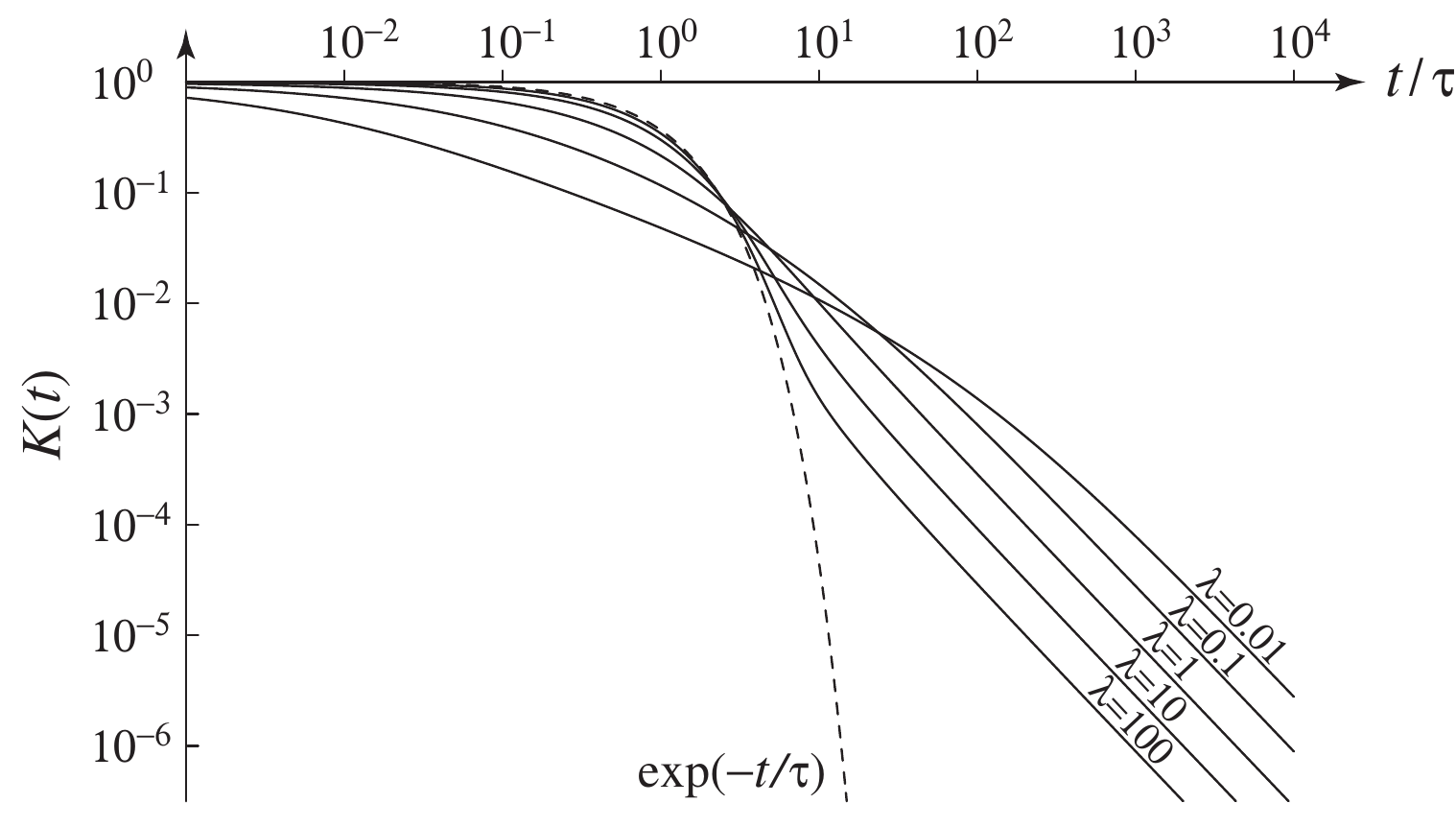}}}
\caption{The function~(\ref{K-int}) versus the rescaled time $t/\tau$ for stick boundary conditions ($b=0$, $\mu=\infty$) and the parameter values $\lambda=0.01$-$100$ (solid lines).  The dashed line shows the exponential decay expected for the standard Langevin process.}
\label{fig3}
\end{figure}

These results are confirmed in Fig.~\ref{fig3}, showing the function $K(t)$ versus time for stick boundary conditions and different values of the parameter $\lambda$.

\subsection{Standard Langevin process}

As already noticed by Lorentz \cite{M02,L21}, the standard Langevin process is recovered in the limit where the fluid has a much lower mass density than the solid particle:
\be
\frac{m_{\rm f}}{m_{\rm s}} = \frac{\rho_{\rm f}}{\rho_{\rm s}} \ll 1 \, .
\label{Lorentz-lim}
\ee
In this case, Eq.~(\ref{Gamma-tilde}) reduces to $\tilde\Gamma(z)=\gamma$ and Eq.~(\ref{K-tilde}) to $\tilde K(z)=(z+1/\tau)^{-1}$.  Therefore, there is a simple pole in the complex plane of the variable $z$ when integrating~(\ref{inv-Laplace}).  We thus find
\bea
&& K(t) = {\rm e}^{-t/\tau} \, , \label{K-Langevin}\\
&& L(t) = \tau \left( 1-{\rm e}^{-t/\tau}\right)  , \\
&& M(t) =\tau t -\tau^2 \left(1- {\rm e}^{-t/\tau}\right)  , 
\eea
so that the classic results of Ref.~\cite{C43} are recovered.

We see in Figs.~\ref{fig2} and~\ref{fig3} that the function $K(t)$ indeed converges towards the form~(\ref{K-Langevin}) in the limit $\lambda\to\infty$ corresponding to the condition~(\ref{Lorentz-lim}).

\section{Long-time random drift and entropy production}
\label{sec:EP}

According to Eq.~(\ref{M-long}), we have that
\be
\lim_{t\to\infty} \frac{1}{t}\, M(t)  = \tau = \frac{m}{\gamma}\, .
\ee
Accordingly, the result~(\ref{p-R}) shows that the particle undergoes a random drift at the asymptotic mean drift velocity
\be
{\bf V}_{\rm drift} = \lim_{t\to\infty}\frac{1}{t}\, \langle {\bf R}(t)\rangle = \frac{1}{\gamma} \, {\bf F}_{\rm ext} \, ,
\ee
and diffusion coefficient
\be
D = \lim_{t\to\infty} \frac{1}{6t} \, \langle\left[{\bf R}(t)-\langle{\bf R}(t)\rangle\right]^2\rangle = \frac{k_{\rm B}T}{\gamma} \, ,
\label{Diff}
\ee
with the friction coefficient~(\ref{friction}) including the effect of the slip length~(\ref{b}).  In the absence of external force, the particle thus performs a random walk of diffusion coefficient~(\ref{Diff}) in any case.  Therefore, the hydrodynamic long-time tail does not modify the Einstein-Sutherland formula for the diffusion coefficient~\cite{E56,S05}.

According to the fluctuation relation~(\ref{FT-R}), the entropy production rate is given by
\be
\frac{1}{k_{\rm B}}\, \frac{d_{\rm i}S}{dt} = \lim_{t\to\infty} \frac{1}{t} \int {\cal P}({\bf R},t\vert{\bf R}_0,0) \ln \frac{{\cal P}({\bf R},t\vert{\bf R}_0,0)}{{\cal P}({\bf R}_0,t\vert{\bf R},0)} \, d{\bf R} = \lim_{t\to\infty} \frac{1}{t} \, \beta \, {\bf F}_{\rm ext}\cdot\langle{\bf R}(t)-{\bf R}(0)\rangle = \beta \, {\bf F}_{\rm ext}\cdot{\bf V}_{\rm drift} \geq 0 \, ,
\ee
in accordance with the second law of thermodynamics.

\section{Concluding remarks}
\label{sec:concl}

In this paper, a nonequilibrium fluctuation theorem is proved within the hydrodynamic theory of Brownian motion.  This theory allows us to obtain the generalized Langevin equation describing the stochastic motion of a colloidal particle in a thermally fluctuating fluid and with slip boundary conditions at the interface between the particle and the fluid.  The stochastic motion is Gaussian, but nonMarkovian because of the hydrodynamic long-time correlations.  Consequently, the velocity autocorrelation function has the well-known $t^{-3/2}$ power-law decay, which is shown to hold for arbitrary values of the slip length.  The early-time decay of the velocity autocorrelation function goes linearly in time for slip boundary conditions, but as the square-root of time for stick boundary conditions.  In the long-time limit, the diffusion coefficient is given by Sutherland's formula for slip boundary conditions~\cite{S05}, which reduces to Einstein's formula for stick boundary conditions~\cite{E56}.

The colloidal particle is considered to be driven away from equilibrium by an external force.  The Gaussian conditional and joint probability distributions for the colloidal particle to evolve in phase space are deduced by the methods of Refs.~\cite{C43,D74}.  The joint probability distributions of some forward and corresponding time-reversed paths are compared.  Remarkably, their ratio only depend on the work performed by the external force on the colloidal particle and the fluid temperature, independently of the hydrodynamic effects.  The fluctuation theorem previously obtained by Kurchan~\cite{K98} for the standard Langevin process is thus shown to hold also for the generalized Langevin process of the hydrodynamic theory of Brownian motion.  In the absence of external force, the principle of detailed balance is recovered, as expected by the microreversibility of the underlying Hamiltonian dynamics of the particle-fluid system.  These results show the generality of the fluctuation theorem to understand the status of the first and second laws of thermodynamics at small scales.

\section*{Acknowledgments}

This paper is dedicated to the memory of Christian Van den Broeck.
The Author thanks Thomas Gilbert, Patrick Grosfils, and Raymond Kapral for useful discussions.
Financial support from the Universit\'e libre de Bruxelles (ULB) and the Fonds de la Recherche Scientifique~-~FNRS under the Grant PDR~T.0094.16 for the project ``SYMSTATPHYS" is acknowledged.

\appendix
\section{Proof of the fluctuation theorem}
\label{appA}

The fluctuation theorem~(\ref{FT}) is proved by direct calculation.  The logarithm of the ratio of the joint probability densities~(\ref{pF}) and~(\ref{pR}) can be expressed as
\be
\ln \frac{p_{\rm F}}{p_{\rm R}} = -\frac{Q_{\rm F}-Q_{\rm R}}{2(FG-H^2)}
\ee
in terms of the quadratic forms associated with the Gaussian distributions:
\bea
&& Q_{\rm F} = G{\bf X}^2-2H{\bf X}\cdot{\bf Y} +F{\bf Y}^2+\beta m (FG-H^2){\bf V}_0^2 \, , \\
&& Q_{\rm R} = G{\bf X'}^2-2H{\bf X'}\cdot{\bf Y'} +F{\bf Y'}^2+\beta m (FG-H^2){\bf V}^2 \, ,
\eea
where ${\bf X}$ and ${\bf Y}$ are given by Eqs.~(\ref{X}) and~(\ref{Y}), while ${\bf X'}$ and ${\bf Y'}$ are obtained from the previous ones by the time-reversal operation giving
\bea
&&{\bf X'} = {\bf R}_0 - {\bf R} + L(t) \, {\bf V} - M(t) \, \frac{{\bf F}_{\rm ext}}{m}\, , \label{X'}\\
&&{\bf Y'} = -{\bf V}_0 + K(t) \, {\bf V} - L(t) \, \frac{{\bf F}_{\rm ext}}{m}\, . \label{Y'}
\eea
Substituting these quantities in the quadratic forms and taking their difference, we get
\bea
Q_{\rm F}-Q_{\rm R} &=& 2\left(GL-H-HK \right) ({\bf R}-{\bf R}_0)\cdot({\bf V}-{\bf V}_0) \nonumber\\
&& - \left[ F(\beta m G - 1 + K^2) +GL^2-2HKL-\beta m H^2\right] ({\bf V}^2-{\bf V}_0^2) \nonumber\\
&& - 2 \left[FL(1-K)-GLM+H(L^2-M+KM)\right] \frac{{\bf F}_{\rm ext}}{m}\cdot({\bf V}+{\bf V}_0) \nonumber\\
&& + 4 \left(HL-GM \right) \frac{{\bf F}_{\rm ext}}{m}\cdot({\bf R}-{\bf R}_0) \, .
\label{DQ1}
\eea
Because of Eqs.~(\ref{F}), (\ref{G}), and~(\ref{H}), we have the identities
\bea
&& GL-H-HK = 0 \, , \\
&& F(\beta m G - 1 + K^2) +GL^2-2HKL-\beta m H^2 = 0 \, , \\
&& FL(1-K)-GLM+H(L^2-M+KM) = 0 \, .
\eea
Accordingly, Eq.~(\ref{DQ1}) reduces to
\be
Q_{\rm F}-Q_{\rm R} = 4 \left(HL-GM \right) \frac{{\bf F}_{\rm ext}}{m}\cdot({\bf R}-{\bf R}_0) \, .
\label{DQ2}
\ee
Moreover, we have that
\be
HL-GM = - \frac{1}{2}\, \beta m \left( FG-H^2\right)  .
\ee
Consequently, we find the identity
\be
\ln \frac{p_{\rm F}}{p_{\rm R}} = \beta\, {\bf F}_{\rm ext}\cdot\left({\bf R}-{\bf R}_0\right),
\ee
so that the fluctuation theorem~(\ref{FT}) is proved.


\section{Analysis of the functions $K(t)$, $L(t)$ and $M(t)$}
\label{appB}

With the new variable $u\equiv rt/\tau$, the integral~(\ref{K-int}) becomes
\be
K(t) = \frac{1}{\pi} \left( \frac{1}{\sqrt{\lambda}}-\frac{1}{\sqrt{\mu}}\right) \left(\frac{\tau}{t}\right)^{3/2} \int_0^{\infty} \frac{\sqrt{u} \ {\rm e}^{-u}}{\left(1-\frac{\tau}{t} u \right)^2 + \frac{\tau}{t} u\left(\frac{1}{\sqrt{\lambda}}-\frac{\tau u}{t\sqrt{\mu}}\right)^2} \, du \, .
\label{K-int2}
\ee
Expanding the integrated function in powers of $\tau/t$, we obtain the asymptotic behavior~(\ref{K-long}).

Next, Eqs.~(\ref{L-long}) and~(\ref{M-long}) are obtained with Eqs.~(\ref{L}) and~(\ref{M}) up to the integration contant $L(\infty)$.  To obtain this latter, we notice that the Laplace transforms of the functions~(\ref{L}) and~(\ref{M}) are given by
\bea
&& \tilde L(z) = \frac{1}{z}\, \tilde K(z) \, , \label{L-tilde}\\
&& \tilde M(z) = \frac{1}{z^2}\, \tilde K(z) \, . \label{M-tilde}
\eea
If we had $L(t)=L(\infty)+O(1/\sqrt{t})$ for $t\to\infty$, its Laplace transform should behave as $\tilde L(z) = L(\infty)/z + O(1/\sqrt{z})$.  Comparing with Eq.~(\ref{L-tilde}), we can thus identify the asymptotic value of the function $L(t)$ as $L(\infty)=\tilde K(0)$.  Therefore, Eqs.~(\ref{K-tilde}) and~(\ref{Gamma-tilde}) imply that $L(\infty)=\tau$, hence the value of the integration constant in Eq.~(\ref{L-long}).  The asymptotic behavior of $M(t)$ is determined consequently.

The short-time behavior of the function $K(t)$ is obtained by expanding $\exp(-rt/\tau)$ in powers of $t/\tau$ in the integral~(\ref{K-int}).  The first two terms of this expansion have finite integrals, so that we get the expression~(\ref{K-short}), but no extra term.

For stick boundary conditions ($b=0$, $\mu=\infty$), even the expression~(\ref{K-short}) is meaningless and another method is required.  Setting $r=x^2/\lambda$ in the integral~(\ref{K-int}) with $\mu=\infty$, we get
\be
K(t) = \frac{1}{\pi} \int_{-\infty}^{+\infty} \frac{x^2 \ {\rm e}^{-\theta^2 x^2}}{(x^2-\lambda)^2 + x^2} \, dx \qquad\mbox{with}\qquad \theta\equiv \sqrt{\frac{t}{\lambda\tau}}\, .
\label{K-int-stick}
\ee
Expressing the Gaussian function $\exp(-\theta^2x^2)$ of $x$ in terms of its Fourier transform and integrating over the variable~$x$ by the method of residues, we find that
\be
K(t) = \frac{2}{\sqrt{\pi}} \int_0^{\infty} {\rm e}^{-q^2} {\rm e}^{-\theta q} \left[ \cos\left( 2\xi \theta q\right) - \frac{1}{2\xi} \,  \sin\left( 2\xi \theta q \right)\right] dq 
\label{K-q}
\ee
with $\xi\equiv \sqrt{\lambda-1/4}$.  Expanding the function in the integral~(\ref{K-q}) in powers of $\theta$, we obtain the result~(\ref{K-short-stick}).



\newpage 

\begin{thebibliography}{51}
\expandafter\ifx\csname natexlab\endcsname\relax\def\natexlab#1{#1}\fi
\expandafter\ifx\csname bibnamefont\endcsname\relax
  \def\bibnamefont#1{#1}\fi
\expandafter\ifx\csname bibfnamefont\endcsname\relax
  \def\bibfnamefont#1{#1}\fi
\expandafter\ifx\csname citenamefont\endcsname\relax
  \def\citenamefont#1{#1}\fi
\expandafter\ifx\csname url\endcsname\relax
  \def\url#1{\texttt{#1}}\fi
\expandafter\ifx\csname urlprefix\endcsname\relax\def\urlprefix{URL }\fi
\providecommand{\bibinfo}[2]{#2}
\providecommand{\eprint}[2][]{\url{#2}}

\bibitem[{\citenamefont{Mazo}(2002)}]{M02}
\bibinfo{author}{\bibfnamefont{R.~M.} \bibnamefont{Mazo}},
  \emph{\bibinfo{title}{Brownian Motion: Fluctuations, Dynamics and
  Applications}} (\bibinfo{publisher}{Clarendon Press},
  \bibinfo{address}{Oxford}, \bibinfo{year}{2002}).

\bibitem[{\citenamefont{Lorentz}(1921)}]{L21}
\bibinfo{author}{\bibfnamefont{H.~A.} \bibnamefont{Lorentz}},
  \emph{\bibinfo{title}{Lessen over theoretische natuurkunde V, Kinetische
  problemen (1911-1912)}} (\bibinfo{publisher}{E. J. Brill},
  \bibinfo{address}{Leiden}, \bibinfo{year}{1921}).

\bibitem[{\citenamefont{Vladimirsky and Terletsky}(1945)}]{VT45}
\bibinfo{author}{\bibfnamefont{V.}~\bibnamefont{Vladimirsky}} \bibnamefont{and}
  \bibinfo{author}{\bibfnamefont{Y.~A.} \bibnamefont{Terletsky}},
  \bibinfo{journal}{Zh. Eksp. Theor. Fiz.} \textbf{\bibinfo{volume}{15}},
  \bibinfo{pages}{258} (\bibinfo{year}{1945}).

\bibitem[{\citenamefont{Dorfman et~al.}(1994)\citenamefont{Dorfman,
  Kirkpatrick, and Sengers}}]{DKS94}
\bibinfo{author}{\bibfnamefont{J.~R.} \bibnamefont{Dorfman}},
  \bibinfo{author}{\bibfnamefont{T.~R.} \bibnamefont{Kirkpatrick}},
  \bibnamefont{and} \bibinfo{author}{\bibfnamefont{J.~V.}
  \bibnamefont{Sengers}}, \bibinfo{journal}{Annu. Rev. Phys. Chem.}
  \textbf{\bibinfo{volume}{45}}, \bibinfo{pages}{213} (\bibinfo{year}{1994}).

\bibitem[{\citenamefont{Alder and Wainwright}(1970)}]{AW70}
\bibinfo{author}{\bibfnamefont{B.~J.} \bibnamefont{Alder}} \bibnamefont{and}
  \bibinfo{author}{\bibfnamefont{T.~E.} \bibnamefont{Wainwright}},
  \bibinfo{journal}{Phys. Rev. A} \textbf{\bibinfo{volume}{1}},
  \bibinfo{pages}{18} (\bibinfo{year}{1970}).

\bibitem[{\citenamefont{Zwanzig and Bixon}(1970)}]{ZB70}
\bibinfo{author}{\bibfnamefont{R.}~\bibnamefont{Zwanzig}} \bibnamefont{and}
  \bibinfo{author}{\bibfnamefont{M.}~\bibnamefont{Bixon}},
  \bibinfo{journal}{Phys. Rev. A} \textbf{\bibinfo{volume}{2}},
  \bibinfo{pages}{2005} (\bibinfo{year}{1970}).

\bibitem[{\citenamefont{Hauge and Martin-L\"of}(1973)}]{HM73}
\bibinfo{author}{\bibfnamefont{E.~H.} \bibnamefont{Hauge}} \bibnamefont{and}
  \bibinfo{author}{\bibfnamefont{A.}~\bibnamefont{Martin-L\"of}},
  \bibinfo{journal}{J. Stat. Phys.} \textbf{\bibinfo{volume}{7}},
  \bibinfo{pages}{259} (\bibinfo{year}{1973}).

\bibitem[{\citenamefont{Dufty}(1974)}]{D74}
\bibinfo{author}{\bibfnamefont{J.~W.} \bibnamefont{Dufty}},
  \bibinfo{journal}{Phys. Fluids} \textbf{\bibinfo{volume}{17}},
  \bibinfo{pages}{328} (\bibinfo{year}{1974}).

\bibitem[{\citenamefont{Clercx and Schram}(1992)}]{CS92}
\bibinfo{author}{\bibfnamefont{H.~J.~H.} \bibnamefont{Clercx}}
  \bibnamefont{and} \bibinfo{author}{\bibfnamefont{P.~P. J.~M.}
  \bibnamefont{Schram}}, \bibinfo{journal}{Phys. Rev. A}
  \textbf{\bibinfo{volume}{46}}, \bibinfo{pages}{1942} (\bibinfo{year}{1992}).

\bibitem[{\citenamefont{Berg-S{\o}rensen and Flyvbjerg}(2005)}]{BF05}
\bibinfo{author}{\bibfnamefont{K.}~\bibnamefont{Berg-S{\o}rensen}}
  \bibnamefont{and}
  \bibinfo{author}{\bibfnamefont{H.}~\bibnamefont{Flyvbjerg}},
  \bibinfo{journal}{New J. Phys.} \textbf{\bibinfo{volume}{7}},
  \bibinfo{pages}{38} (\bibinfo{year}{2005}).

\bibitem[{\citenamefont{Paul and Pusey}(1981)}]{PP81}
\bibinfo{author}{\bibfnamefont{G.~L.} \bibnamefont{Paul}} \bibnamefont{and}
  \bibinfo{author}{\bibfnamefont{P.~N.} \bibnamefont{Pusey}},
  \bibinfo{journal}{J. Phys. A: Math. Gen.} \textbf{\bibinfo{volume}{14}},
  \bibinfo{pages}{3301} (\bibinfo{year}{1981}).

\bibitem[{\citenamefont{Jeney et~al.}(2008)\citenamefont{Jeney, Luki\'c, Kraus,
  Franosch, and Forr\'o}}]{JLKFF08}
\bibinfo{author}{\bibfnamefont{S.}~\bibnamefont{Jeney}},
  \bibinfo{author}{\bibfnamefont{B.}~\bibnamefont{Luki\'c}},
  \bibinfo{author}{\bibfnamefont{J.~A.} \bibnamefont{Kraus}},
  \bibinfo{author}{\bibfnamefont{T.}~\bibnamefont{Franosch}}, \bibnamefont{and}
  \bibinfo{author}{\bibfnamefont{L.}~\bibnamefont{Forr\'o}},
  \bibinfo{journal}{Phys. Rev. Lett.} \textbf{\bibinfo{volume}{100}},
  \bibinfo{pages}{240604} (\bibinfo{year}{2008}).

\bibitem[{\citenamefont{Franosch et~al.}(2011)\citenamefont{Franosch, Grimm,
  Belushkin, Mor, Foffi, Forr\'o, and Jeney}}]{FGBMFFJ11}
\bibinfo{author}{\bibfnamefont{T.}~\bibnamefont{Franosch}},
  \bibinfo{author}{\bibfnamefont{M.}~\bibnamefont{Grimm}},
  \bibinfo{author}{\bibfnamefont{M.}~\bibnamefont{Belushkin}},
  \bibinfo{author}{\bibfnamefont{F.~M.} \bibnamefont{Mor}},
  \bibinfo{author}{\bibfnamefont{G.}~\bibnamefont{Foffi}},
  \bibinfo{author}{\bibfnamefont{L.}~\bibnamefont{Forr\'o}}, \bibnamefont{and}
  \bibinfo{author}{\bibfnamefont{S.}~\bibnamefont{Jeney}},
  \bibinfo{journal}{Nature} \textbf{\bibinfo{volume}{478}}, \bibinfo{pages}{85}
  (\bibinfo{year}{2011}).

\bibitem[{\citenamefont{Kheifets et~al.}(2014)\citenamefont{Kheifets, Simha,
  Melin, Li, and Raizen}}]{KSMLR14}
\bibinfo{author}{\bibfnamefont{S.}~\bibnamefont{Kheifets}},
  \bibinfo{author}{\bibfnamefont{A.}~\bibnamefont{Simha}},
  \bibinfo{author}{\bibfnamefont{K.}~\bibnamefont{Melin}},
  \bibinfo{author}{\bibfnamefont{T.}~\bibnamefont{Li}}, \bibnamefont{and}
  \bibinfo{author}{\bibfnamefont{M.~G.} \bibnamefont{Raizen}},
  \bibinfo{journal}{Science} \textbf{\bibinfo{volume}{343}},
  \bibinfo{pages}{1493} (\bibinfo{year}{2014}).

\bibitem[{\citenamefont{Langevin}(1908)}]{L08}
\bibinfo{author}{\bibfnamefont{P.}~\bibnamefont{Langevin}},
  \bibinfo{journal}{C. R. Acad. Sci. (Paris)} \textbf{\bibinfo{volume}{146}},
  \bibinfo{pages}{530} (\bibinfo{year}{1908}).

\bibitem[{\citenamefont{Bedeaux and Mazur}(1974{\natexlab{a}})}]{BM74}
\bibinfo{author}{\bibfnamefont{D.}~\bibnamefont{Bedeaux}} \bibnamefont{and}
  \bibinfo{author}{\bibfnamefont{P.}~\bibnamefont{Mazur}},
  \bibinfo{journal}{Physica A} \textbf{\bibinfo{volume}{76}},
  \bibinfo{pages}{247} (\bibinfo{year}{1974}{\natexlab{a}}).

\bibitem[{\citenamefont{Albano et~al.}(1975)\citenamefont{Albano, Bedeaux, and
  Mazur}}]{ABM75}
\bibinfo{author}{\bibfnamefont{A.~M.} \bibnamefont{Albano}},
  \bibinfo{author}{\bibfnamefont{D.}~\bibnamefont{Bedeaux}}, \bibnamefont{and}
  \bibinfo{author}{\bibfnamefont{P.}~\bibnamefont{Mazur}},
  \bibinfo{journal}{Physica A} \textbf{\bibinfo{volume}{80}},
  \bibinfo{pages}{89} (\bibinfo{year}{1975}).

\bibitem[{\citenamefont{Bedeaux et~al.}(1977)\citenamefont{Bedeaux, Albano, and
  Mazur}}]{BAM77}
\bibinfo{author}{\bibfnamefont{D.}~\bibnamefont{Bedeaux}},
  \bibinfo{author}{\bibfnamefont{A.~M.} \bibnamefont{Albano}},
  \bibnamefont{and} \bibinfo{author}{\bibfnamefont{P.}~\bibnamefont{Mazur}},
  \bibinfo{journal}{Physica A} \textbf{\bibinfo{volume}{88}},
  \bibinfo{pages}{574} (\bibinfo{year}{1977}).

\bibitem[{\citenamefont{Felderhof}(1976{\natexlab{a}})}]{F76a}
\bibinfo{author}{\bibfnamefont{B.~U.} \bibnamefont{Felderhof}},
  \bibinfo{journal}{Physica A} \textbf{\bibinfo{volume}{84}},
  \bibinfo{pages}{557} (\bibinfo{year}{1976}{\natexlab{a}}).

\bibitem[{\citenamefont{Felderhof}(1976{\natexlab{b}})}]{F76b}
\bibinfo{author}{\bibfnamefont{B.~U.} \bibnamefont{Felderhof}},
  \bibinfo{journal}{Physica A} \textbf{\bibinfo{volume}{84}},
  \bibinfo{pages}{569} (\bibinfo{year}{1976}{\natexlab{b}}).

\bibitem[{\citenamefont{Felderhof}(1977)}]{F77}
\bibinfo{author}{\bibfnamefont{B.~U.} \bibnamefont{Felderhof}},
  \bibinfo{journal}{Physica A} \textbf{\bibinfo{volume}{88}},
  \bibinfo{pages}{614} (\bibinfo{year}{1977}).

\bibitem[{\citenamefont{Kramers}(1940)}]{K40}
\bibinfo{author}{\bibfnamefont{H.~A.} \bibnamefont{Kramers}},
  \bibinfo{journal}{Physica} \textbf{\bibinfo{volume}{7}}, \bibinfo{pages}{284}
  (\bibinfo{year}{1940}).

\bibitem[{\citenamefont{Chandrasekhar}(1943)}]{C43}
\bibinfo{author}{\bibfnamefont{S.}~\bibnamefont{Chandrasekhar}},
  \bibinfo{journal}{Rev. Mod. Phys.} \textbf{\bibinfo{volume}{15}},
  \bibinfo{pages}{1} (\bibinfo{year}{1943}).

\bibitem[{\citenamefont{Kurchan}(1998)}]{K98}
\bibinfo{author}{\bibfnamefont{J.}~\bibnamefont{Kurchan}}, \bibinfo{journal}{J.
  Phys. A: Math. Gen.} \textbf{\bibinfo{volume}{31}}, \bibinfo{pages}{3719}
  (\bibinfo{year}{1998}).

\bibitem[{\citenamefont{Crooks}(1999)}]{C99}
\bibinfo{author}{\bibfnamefont{G.~E.} \bibnamefont{Crooks}},
  \bibinfo{journal}{Phys. Rev. E} \textbf{\bibinfo{volume}{60}},
  \bibinfo{pages}{2721} (\bibinfo{year}{1999}).

\bibitem[{\citenamefont{Jarzynski}(2011)}]{J11}
\bibinfo{author}{\bibfnamefont{C.}~\bibnamefont{Jarzynski}},
  \bibinfo{journal}{Annu. Rev. Condens. Matter Phys.}
  \textbf{\bibinfo{volume}{2}}, \bibinfo{pages}{329} (\bibinfo{year}{2011}).

\bibitem[{\citenamefont{Blickle et~al.}(2006)\citenamefont{Blickle, Speck,
  Helden, Seifert, and Bechinger}}]{BSHSB06}
\bibinfo{author}{\bibfnamefont{V.}~\bibnamefont{Blickle}},
  \bibinfo{author}{\bibfnamefont{T.}~\bibnamefont{Speck}},
  \bibinfo{author}{\bibfnamefont{L.}~\bibnamefont{Helden}},
  \bibinfo{author}{\bibfnamefont{U.}~\bibnamefont{Seifert}}, \bibnamefont{and}
  \bibinfo{author}{\bibfnamefont{C.}~\bibnamefont{Bechinger}},
  \bibinfo{journal}{Phys. Rev. Lett.} \textbf{\bibinfo{volume}{96}},
  \bibinfo{pages}{070603} (\bibinfo{year}{2006}).

\bibitem[{\citenamefont{Ciliberto et~al.}(2013)\citenamefont{Ciliberto,
  Gomez-Solano, and Petrosyan}}]{CGP13}
\bibinfo{author}{\bibfnamefont{S.}~\bibnamefont{Ciliberto}},
  \bibinfo{author}{\bibfnamefont{R.}~\bibnamefont{Gomez-Solano}},
  \bibnamefont{and}
  \bibinfo{author}{\bibfnamefont{A.}~\bibnamefont{Petrosyan}},
  \bibinfo{journal}{Annu. Rev. Condens. Matter Phys.}
  \textbf{\bibinfo{volume}{4}}, \bibinfo{pages}{235} (\bibinfo{year}{2013}).

\bibitem[{\citenamefont{Zamponi et~al.}(2005)\citenamefont{Zamponi, Bonetto,
  Cugliandolo, and Kurchan}}]{ZBCK05}
\bibinfo{author}{\bibfnamefont{F.}~\bibnamefont{Zamponi}},
  \bibinfo{author}{\bibfnamefont{F.}~\bibnamefont{Bonetto}},
  \bibinfo{author}{\bibfnamefont{L.~F.} \bibnamefont{Cugliandolo}},
  \bibnamefont{and} \bibinfo{author}{\bibfnamefont{J.}~\bibnamefont{Kurchan}},
  \bibinfo{journal}{J. Stat. Mech.: Th. Exp.} \textbf{\bibinfo{volume}{2005}},
  \bibinfo{pages}{P09013} (\bibinfo{year}{2005}).

\bibitem[{\citenamefont{Speck and Seifert}(2007)}]{SS07}
\bibinfo{author}{\bibfnamefont{T.}~\bibnamefont{Speck}} \bibnamefont{and}
  \bibinfo{author}{\bibfnamefont{U.}~\bibnamefont{Seifert}},
  \bibinfo{journal}{J. Stat. Mech.: Th. Exp.} \textbf{\bibinfo{volume}{2007}},
  \bibinfo{pages}{L09002} (\bibinfo{year}{2007}).

\bibitem[{\citenamefont{Mai and Dhar}(2007)}]{MD07}
\bibinfo{author}{\bibfnamefont{T.}~\bibnamefont{Mai}} \bibnamefont{and}
  \bibinfo{author}{\bibfnamefont{A.}~\bibnamefont{Dhar}},
  \bibinfo{journal}{Phys. Rev. E} \textbf{\bibinfo{volume}{75}},
  \bibinfo{pages}{061101} (\bibinfo{year}{2007}).

\bibitem[{\citenamefont{Esposito and Lindenberg}(2008)}]{EL08}
\bibinfo{author}{\bibfnamefont{M.}~\bibnamefont{Esposito}} \bibnamefont{and}
  \bibinfo{author}{\bibfnamefont{K.}~\bibnamefont{Lindenberg}},
  \bibinfo{journal}{Phys. Rev. E} \textbf{\bibinfo{volume}{77}},
  \bibinfo{pages}{051119} (\bibinfo{year}{2008}).

\bibitem[{\citenamefont{Andrieux and Gaspard}(2008)}]{AG08JSM}
\bibinfo{author}{\bibfnamefont{D.}~\bibnamefont{Andrieux}} \bibnamefont{and}
  \bibinfo{author}{\bibfnamefont{P.}~\bibnamefont{Gaspard}},
  \bibinfo{journal}{J. Stat. Mech.: Th. Exp.} \textbf{\bibinfo{volume}{2008}},
  \bibinfo{pages}{P11007} (\bibinfo{year}{2008}).

\bibitem[{\citenamefont{Puglisi and Villamaina}(2009)}]{PV09}
\bibinfo{author}{\bibfnamefont{A.}~\bibnamefont{Puglisi}} \bibnamefont{and}
  \bibinfo{author}{\bibfnamefont{D.}~\bibnamefont{Villamaina}},
  \bibinfo{journal}{EPL} \textbf{\bibinfo{volume}{88}}, \bibinfo{pages}{30004}
  (\bibinfo{year}{2009}).

\bibitem[{\citenamefont{Aron et~al.}(2010)\citenamefont{Aron, Biroli, and
  Cugliandolo}}]{ABC10}
\bibinfo{author}{\bibfnamefont{C.}~\bibnamefont{Aron}},
  \bibinfo{author}{\bibfnamefont{G.}~\bibnamefont{Biroli}}, \bibnamefont{and}
  \bibinfo{author}{\bibfnamefont{L.~F.} \bibnamefont{Cugliandolo}},
  \bibinfo{journal}{J. Stat. Mech.: Th. Exp.} \textbf{\bibinfo{volume}{2010}},
  \bibinfo{pages}{P11018} (\bibinfo{year}{2010}).

\bibitem[{\citenamefont{Maes et~al.}(2013)\citenamefont{Maes, Safaverdi, Visco,
  and van Wijland}}]{MSVvW13}
\bibinfo{author}{\bibfnamefont{C.}~\bibnamefont{Maes}},
  \bibinfo{author}{\bibfnamefont{S.}~\bibnamefont{Safaverdi}},
  \bibinfo{author}{\bibfnamefont{P.}~\bibnamefont{Visco}}, \bibnamefont{and}
  \bibinfo{author}{\bibfnamefont{F.}~\bibnamefont{van Wijland}},
  \bibinfo{journal}{Phys. Rev. E} \textbf{\bibinfo{volume}{87}},
  \bibinfo{pages}{022125} (\bibinfo{year}{2013}).

\bibitem[{\citenamefont{Chechkin and Klages}(2009)}]{CK09}
\bibinfo{author}{\bibfnamefont{A.~V.} \bibnamefont{Chechkin}} \bibnamefont{and}
  \bibinfo{author}{\bibfnamefont{R.}~\bibnamefont{Klages}},
  \bibinfo{journal}{J. Stat. Mech.: Th. Exp.} \textbf{\bibinfo{volume}{2009}},
  \bibinfo{pages}{L03002} (\bibinfo{year}{2009}).

\bibitem[{\citenamefont{Chechkin et~al.}(2012)\citenamefont{Chechkin, Lenz, and
  Klages}}]{CLK12}
\bibinfo{author}{\bibfnamefont{A.~V.} \bibnamefont{Chechkin}},
  \bibinfo{author}{\bibfnamefont{F.}~\bibnamefont{Lenz}}, \bibnamefont{and}
  \bibinfo{author}{\bibfnamefont{R.}~\bibnamefont{Klages}},
  \bibinfo{journal}{J. Stat. Mech.: Th. Exp.} \textbf{\bibinfo{volume}{2012}},
  \bibinfo{pages}{L11001} (\bibinfo{year}{2012}).

\bibitem[{\citenamefont{Dieterich et~al.}(2015)\citenamefont{Dieterich, Klages,
  and Chechkin}}]{DKC15}
\bibinfo{author}{\bibfnamefont{P.}~\bibnamefont{Dieterich}},
  \bibinfo{author}{\bibfnamefont{R.}~\bibnamefont{Klages}}, \bibnamefont{and}
  \bibinfo{author}{\bibfnamefont{A.~V.} \bibnamefont{Chechkin}},
  \bibinfo{journal}{New J. Phys.} \textbf{\bibinfo{volume}{17}},
  \bibinfo{pages}{075004} (\bibinfo{year}{2015}).

\bibitem[{\citenamefont{Landau and Lifshitz}(1980)}]{LL80Part2}
\bibinfo{author}{\bibfnamefont{L.~D.} \bibnamefont{Landau}} \bibnamefont{and}
  \bibinfo{author}{\bibfnamefont{E.~M.} \bibnamefont{Lifshitz}},
  \emph{\bibinfo{title}{Statistical Physics, Part~2}}
  (\bibinfo{publisher}{Pergamon Press}, \bibinfo{address}{Oxford},
  \bibinfo{year}{1980}).

\bibitem[{\citenamefont{{Ortiz~de~Z\'arate} and Sengers}(2006)}]{OS06}
\bibinfo{author}{\bibfnamefont{J.~M.} \bibnamefont{{Ortiz~de~Z\'arate}}}
  \bibnamefont{and} \bibinfo{author}{\bibfnamefont{J.~V.}
  \bibnamefont{Sengers}}, \emph{\bibinfo{title}{Hydrodynamic Fluctuations in
  Fluids and Fluid Mixtures}} (\bibinfo{publisher}{Elsevier},
  \bibinfo{address}{Amsterdam}, \bibinfo{year}{2006}).

\bibitem[{\citenamefont{Bedeaux and Mazur}(1974{\natexlab{b}})}]{BM74c}
\bibinfo{author}{\bibfnamefont{D.}~\bibnamefont{Bedeaux}} \bibnamefont{and}
  \bibinfo{author}{\bibfnamefont{P.}~\bibnamefont{Mazur}},
  \bibinfo{journal}{Physica A} \textbf{\bibinfo{volume}{78}},
  \bibinfo{pages}{505} (\bibinfo{year}{1974}{\natexlab{b}}).

\bibitem[{\citenamefont{Mazur and Bedeaux}(1974)}]{MB74}
\bibinfo{author}{\bibfnamefont{P.}~\bibnamefont{Mazur}} \bibnamefont{and}
  \bibinfo{author}{\bibfnamefont{D.}~\bibnamefont{Bedeaux}},
  \bibinfo{journal}{Physica A} \textbf{\bibinfo{volume}{76}},
  \bibinfo{pages}{235} (\bibinfo{year}{1974}).

\bibitem[{\citenamefont{Bedeaux et~al.}(1976)\citenamefont{Bedeaux, Albano, and
  Mazur}}]{BAM76}
\bibinfo{author}{\bibfnamefont{D.}~\bibnamefont{Bedeaux}},
  \bibinfo{author}{\bibfnamefont{A.~M.} \bibnamefont{Albano}},
  \bibnamefont{and} \bibinfo{author}{\bibfnamefont{P.}~\bibnamefont{Mazur}},
  \bibinfo{journal}{Physica A} \textbf{\bibinfo{volume}{82}},
  \bibinfo{pages}{438} (\bibinfo{year}{1976}).

\bibitem[{\citenamefont{Gaspard and Kapral}(2018)}]{GK18a}
\bibinfo{author}{\bibfnamefont{P.}~\bibnamefont{Gaspard}} \bibnamefont{and}
  \bibinfo{author}{\bibfnamefont{R.}~\bibnamefont{Kapral}},
  \bibinfo{journal}{J. Chem. Phys.} \textbf{\bibinfo{volume}{148}},
  \bibinfo{pages}{134104} (\bibinfo{year}{2018}).

\bibitem[{\citenamefont{Corrsin and Lumley}(1956)}]{CL56}
\bibinfo{author}{\bibfnamefont{S.}~\bibnamefont{Corrsin}} \bibnamefont{and}
  \bibinfo{author}{\bibfnamefont{J.}~\bibnamefont{Lumley}},
  \bibinfo{journal}{Appl. Sci. Res. A} \textbf{\bibinfo{volume}{6}},
  \bibinfo{pages}{114} (\bibinfo{year}{1956}).

\bibitem[{\citenamefont{Maxey and Riley}(1983)}]{MR83}
\bibinfo{author}{\bibfnamefont{M.~R.} \bibnamefont{Maxey}} \bibnamefont{and}
  \bibinfo{author}{\bibfnamefont{J.~J.} \bibnamefont{Riley}},
  \bibinfo{journal}{Phys. Fluids} \textbf{\bibinfo{volume}{26}},
  \bibinfo{pages}{883} (\bibinfo{year}{1983}).

\bibitem[{\citenamefont{Balakrishnan}(1979)}]{B79}
\bibinfo{author}{\bibfnamefont{V.}~\bibnamefont{Balakrishnan}},
  \bibinfo{journal}{Pramana} \textbf{\bibinfo{volume}{12}},
  \bibinfo{pages}{301} (\bibinfo{year}{1979}).

\bibitem[{\citenamefont{Sekimoto}(2010)}]{S10}
\bibinfo{author}{\bibfnamefont{K.}~\bibnamefont{Sekimoto}},
  \emph{\bibinfo{title}{Stochastic Energetics, Lect. Notes Phys. 799}}
  (\bibinfo{publisher}{Springer}, \bibinfo{address}{Berlin},
  \bibinfo{year}{2010}).

\bibitem[{\citenamefont{Einstein}(1956)}]{E56}
\bibinfo{author}{\bibfnamefont{A.}~\bibnamefont{Einstein}},
  \emph{\bibinfo{title}{Investigations on the theory of the {Brownian}
  movement}} (\bibinfo{publisher}{Dover}, \bibinfo{address}{New York},
  \bibinfo{year}{1956}).

\bibitem[{\citenamefont{Sutherland}(1905)}]{S05}
\bibinfo{author}{\bibfnamefont{W.}~\bibnamefont{Sutherland}},
  \bibinfo{journal}{Phil. Mag.} \textbf{\bibinfo{volume}{9}},
  \bibinfo{pages}{781} (\bibinfo{year}{1905}).

\end{thebibliography}
\end{document}